\documentstyle[aps,prl,floats,psfig]{revtex}

\begin{document}

\draft
\wideabs{

\title{Anisotropic, Fermi-Surface-Induced Variation in \protect\( T_{c}\protect \)
in \protect\( MgB_{2}\protect \) Alloys }

\author{Prabhakar P. Singh}

\address{Department of Physics, Indian Institute of Technology, Powai, Mumbai- 400076,
India}
\date{\today}
\maketitle

\begin{abstract}
Using coherent-potential to describe 
disorder, Gaspari-Gyorffy approach to evaluate  electron-phonon coupling, 
and Allen-Dynes equation to calculate  $T_{c}$,
we show that in $Mg_{1-x}M_{x}B_{2}$ ($M\equiv Al,Li$ or $Zn$)
alloys (i) the way $T_{c}$ changes depends on the location of the added/modified
{\bf k}-resolved states on the Fermi surface and (ii) the variation of 
$T_{c}$
 as a function
of concentration is dictated by the $B$ $p$ DOS.
 In addition, using full-potential 
 calculations for  $MgMB_{4}$, 
we show that (i) at $x=0.5$ a superstructure
can form in $Mg_{1-x}Al_{x}B_{2}$ but not in $Mg_{1-x}Li_{x}B_{2}$
or $Mg_{1-x}Zn_{x}B_{2}$, and (ii) $B$ layer shifts towards the
impurity layer, more for $Al$ than for  $Li$ or $Zn$.
\end{abstract}

 } 

Since the discovery of superconductivity in \( MgB_{2} \) \cite{nag} the experimental
\cite{nag,bud,hin,tak,rev,yil} and theoretical 
\cite{kor,an1,kon,boh,pps1,med,sat,bel,liu,cho} efforts have greatly improved our understanding
of the nature of interaction responsible for superconductivity (SC) in \( MgB_{2}. \) It has become
clear that almost all facets of the phonon-mediated electron-electron interaction
have a dramatic influence over the superconducting behavior of \( MgB_{2}. \)
For example, the electron-phonon matrix elements change considerably as one
moves away from the cylindrical Fermi sheets along \( \Gamma  \) to \( A \)
\cite{kon,cho}, anharmonic effects \cite{yil,liu} have to be included in the dynamical
matrix \cite{cho}, and finally {\bf k}-dependent fully anisotropic Eliashberg
equations have to be solved \cite{cho} for a complete and accurate description of the superconducting
properties  of \( MgB_{2}. \) Such a strong dependence of the superconducting properties
of \( MgB_{2} \) on different aspects of the interaction has opened up the
possibility of dramatically modifying its superconducting behavior by changing
the interaction in various ways and thereby learning more about the interaction
itself. Alloying \( MgB_{2} \) with various elements and then studying their
SC properties offers such an opportunity.

There have been several studies of changes in 
the SC properties of \( MgB_{2} \)
upon substitutions of various elements such as \( Be,\, Li,\, C,\, Al,\, Na,\, Zn,\, Zr,\,  \)\( Fe,\, Co,\, Ni, \)
and others \cite{rev,xia,zha,kaz,mor}. The main effects of alloying are seen to be (i) a decrease in transition
temperature, \( T_{c} \), with increasing concentration of the alloying elements
although the rate at which the \( T_{c} \) changes depends on the element being
substituted, (ii) a slight increase in the \( T_{c} \) in case of \( Zn \)
 \cite{kaz,mor} 
substitution while for \( Si \) and \( Li \) the \( T_{c} \) remains essentially
the same, (iii) persistence of superconductivity up to \( x\sim 0.7 \) in \( Mg_{1-x}Al_{x}B_{2} \) \cite{rev,xia}, 
(iv) a change in crystal structure and the formation of a superstructure at
\( x=0.5 \) in \( Mg_{1-x}Al_{x}B_{2} \) \cite{xia}, and (v) a change in the lattice
parameters \( a \) and \( c. \) 

In an effort to understand the changes in the electronic structure and the superconducting
properties of \( MgB_{2} \) alloys, we have carried out {\it ab  initio }
studies of \( Mg_{1-x}Al_{x}B_{2}, \) \( Mg_{1-x}Li_{x}B_{2} \) and \( Mg_{1-x}Zn_{x}B_{2} \)
alloys. We have used Korringa-Kohn-Rostoker coherent-potential approximation
\cite{pps_cpa,fau} in the atomic-sphere approximation (KKR-ASA CPA) method for taking
into account the effects of disorder, Gaspari-Gyorffy formalism \cite{gas} for calculating
the electron-phonon coupling constant \( \lambda  \), and Allen-Dynes equation 
\cite{all} 
for calculating \( T_{c} \) in \( Mg_{1-x}Al_{x}B_{2}, \) \( Mg_{1-x}Li_{x}B_{2} \)
and \( Mg_{1-x}Zn_{x}B_{2} \) alloys as a function of \( Al,\, Li \) and \( Zn \)
concentrations, respectively. We have analyzed our results in terms of the changes
in the spectral function \cite{fau} along \( \Gamma  \)  to \( A \) evaluated 
at the Fermi energy, $E_F$, and the total density
of states (DOS), in particular the changes in the \( B \) \( p \) contribution
to the total DOS, as a function of concentration \( x \). 

For examining the possibility of superstructure formation at \( x=0.5, \) we
have used ABINIT code \cite{abinit}, based on psuedopotentials and plane waves to optimize
the cell parameters \( a \) and \( c \) as well as relax the cell-internal
atomic positions of \( MgAlB_{4}, \) \( MgLiB_{4} \) and \( MgZnB_{4} \)
in \( P6/mmm \) structures. We have used these atomic positions to carry out
a total energy comparison using KKR-ASA CPA between the ordered and the substitutionally
disordered \( Mg_{1-x}Al_{x}B_{2},\,  \)\( Mg_{1-x}Li_{x}B_{2} \) and \( Mg_{1-x}Zn_{x}B_{2} \)
 alloys at \( x=0.5 \). Such an approach allows us to check 
the possibility of formation of a layered or a mixed superstructure at \( x=0.5 \)
in these alloys. Before we describe our results, we outline some of the computational
details.
\begin{figure}
\centering
\psfig{file=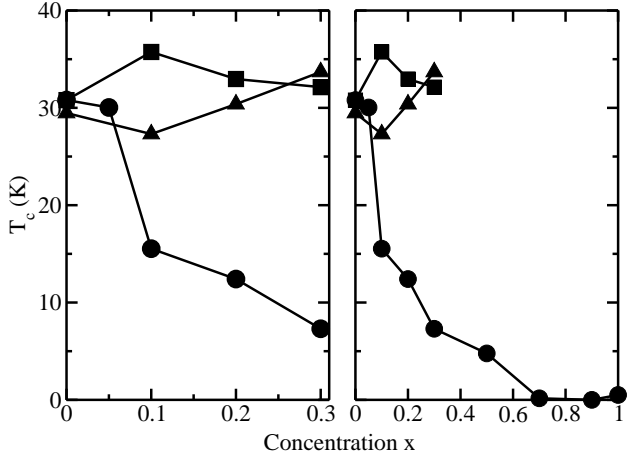,height=7.4cm,angle=-90}
\vspace{0.3cm}
\caption{The calculated variation of $T_c$ as a function of concentration $x$ in
 $Mg_{1-x}Al_xB_2$, 
 $Mg_{1-x}Li_xB_2$ and 
 $Mg_{1-x}Zn_xB_2$ alloys. 
}
\end{figure}

The charge self-consistent electronic structure of \( Mg_{1-x}Al_{x}B_{2}, \)
\( Mg_{1-x}Li_{x}B_{2} \) and \( Mg_{1-x}Zn_{x}B_{2} \) alloys as a function
of \( x \) has been calculated using the KKR-ASA CPA method. We have
used the CPA rather than a rigid-band model because CPA has been found to reliably
describe the effects of disorder in metallic alloys 
\cite{pps_cpa,fau}. We parametrized the exchange-correlation
potential as suggested by Perdew-Wang \cite{perdew} within the generalized gradient
approximation. The Brillouin zone (BZ) integration was carried out using \( 1215 \)
\( k- \) points in the irreducible part of the BZ. For DOS and 
spectral function calculations, we
added a small imaginary component of \( 1 \) \( mRy \) 
and $2$  $mRy$, respectively, to the energy and used
\( 4900 \) {\bf k}-points in the irreducible part of the BZ. The lattice
constants for \( Mg_{1-x}Al_{x}B_{2}, \) \( Mg_{1-x}Li_{x}B_{2} \) and \( Mg_{1-x}Zn_{x}B_{2} \)
alloys as a function of \( x \) were taken from experiments 
\cite{xia,zha,kaz}. The Wigner- Seitz
radii for \( Mg \),   \( Al \) and \( Zn \) were slightly larger than that
of \( B \). The sphere overlap which is crucial in ASA, was less than \( 10 \)\%
and the maximum \( l \) used was \( l_{max} \) = \( 3 \).

The electron-phonon coupling constant \( \lambda  \) was calculated using Gaspari-Gyorffy
\cite{gas} formalism with the charge self-consistent potentials of \( Mg_{1-x}Al_{x}B_{2}, \)
\( Mg_{1-x}Li_{x}B_{2} \) and \( Mg_{1-x}Zn_{x}B_{2} \) obtained with the
KKR-ASA CPA method. Subsequently, the variation of \( T_{c} \) as a function
of \( Al \), \( Li \) and \( Zn \) concentrations was calculated using Allen-Dynes
equation \cite{all}. The average values of phonon frequencies \( \omega _{ln} \)
for \( MgB_{2} \) and \( AlB_{2} \) were taken from Refs. \cite{kon,boh} respectively.
For intermediate concentrations, we took \( \omega _{ln} \) to be the concentration-weighted
average of \( MgB_{2} \) and \( AlB_{2} \). For \( Mg_{1-x}Li_{x}B_{2} \)
and \( Mg_{1-x}Zn_{x}B_{2} \) we used the same value of \( \omega _{ln} \)
as that for \( MgB_{2} \). 

The structural relaxation of \( MgALB_{4}, \) \( MgLiB_{4} \) and \( MgZnB_{4} \)
was carried out by the molecular dynamics program ABINIT with Broyden-Fletcher-Goldfarb-Shanno
minimization technique \cite{abinit} using Troullier-Martins psuedopotentials \cite{tro}, 512 Monkhorst-Pack \cite{monkhorst} 
{\bf k}-points and Teter parameterization \cite{abinit} for exchange-correlation. The kinetic energy
cutoff for the plane waves was 110 Ry. 

Based on our calculations, described below, we find that in \( Mg_{1-x}Al_{x}B_{2}, \)
\( Mg_{1-x}Li_{x}B_{2} \) and \( Mg_{1-x}Zn_{x}B_{2} \) alloys (i) the way
\( T_{c} \) changes depends on the location of the added/modified \textbf{k}-resolved
states on the Fermi surface, (ii) the variation of \( T_{c} \) as a function
of concentration is dictated by the \( B\, p \) contribution to the total DOS,
(iii) at \( x=0.5 \) a superstructure can form in \( Mg_{1-x}Al_{x}B_{2} \)
but not in \( Mg_{1-x}Li_{x}B_{2} \) or \( Mg_{1-x}Zn_{x}B_{2} \), and (iv)
\( B \) layer shifts towards the impurity layer, more for \( Al \) than
for \( Li \) or \( Zn. \)
\begin{figure}
\centering
\psfig{file=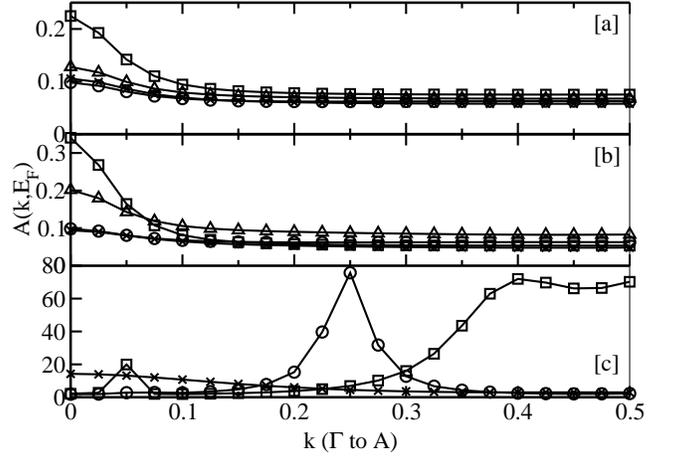,height=7.5cm,angle=-90}
\vspace{0.3cm}
\caption{The calculated spectral function along $\Gamma$ to $A$, evaluated at the Fermi energy, as a function of concentration $x$ in 
 $Mg_{1-x}Al_xB_2$, 
 $Mg_{1-x}Li_xB_2$ and 
 $Mg_{1-x}Zn_xB_2$ alloys. Figures (a)  and (b) correspond to $x=0.1$ and 
$x=0.3$, respectively and the symbols open circle, open square, x and open 
triangle correspond to $MgB_2$, 
 $Mg_{1-x}Al_xB_2$, 
 $Mg_{1-x}Li_xB_2$ and 
 $Mg_{1-x}Zn_xB_2$ alloys respectively. In figure (c)  the symbols open circle, open square 
and x correspond to $AlB_2$, 
 $Mg_{0.1}Al_{0.9}B_2$ and 
 $Mg_{0.4}Al_{0.6}B_2$, respectively. For clarity, in figure (c) we have multiplied the spectral function  of $Mg_{0.4}Al_{0.6}B_2$ by 100.
}
\end{figure}

The main results of our calculations are shown in Fig. 1, where we have plotted
the variation in \( T_{c} \) of \( Mg_{1-x}Al_{x}B_{2}, \) \( Mg_{1-x}Li_{x}B_{2} \)
and \( Mg_{1-x}Zn_{x}B_{2} \) alloys as a function of concentration \( x \).
The calculations were carried out as described earlier with the same value of
\( \mu ^{*}=0.09 \) for all the concentrations. The \( T_{c} \) for \( MgB_{2} \)
is equal to \( \sim 30.8\, K, \) which is consistent with the results of other
works \cite{kor,kon,boh} with similar approximations. The corresponding \( \lambda  \)
is equal to \( 0.69. \) For \( 0<x\leq 0.3 \), the \( T_{c} \) increases
slightly for \( Mg_{1-x}Li_{x}B_{2} \) and \( Mg_{1-x}Zn_{x}B_{2} \), while
it decreases substantially for \( Mg_{1-x}Al_{x}B_{2}, \) as is found experimentally \cite{rev,xia}.
Note that for \( x=0.1, \) our calculation shows \( Mg_{1-x}Li_{x}B_{2} \)
to have a \( T_{c} \) higher than that of \( Mg_{1-x}Zn_{x}B_{2} \) by about
\( 7\, K \) \cite{zha,kaz,mor}. In Fig.1 (right panel) we have shown the variation in \( T_{c} \)
in \( Mg_{1-x}Al_{x}B_{2} \) alloys as a function of concentration for \( 0\leq x\leq 1. \)
As a function of \( Al \) concentration, the \( T_{c} \) decrease rapidly
from \( 30\, K \) at \( x=0.05 \) to about \( 15\, K \) at \( x=0.1. \)
The \( T_{c} \) decreases slowly between \( x=0.4 \) and \( x=0.5. \) At
\( x=0.7 \) the \( T_{c} \) vanishes and remains essentially zero thereafter.
The calculated variation in \( T_{c} \), as shown in Fig. 1, is in very good
qualitative agreement with the experiments \cite{rev}. 
\begin{figure}
\centering
\psfig{file=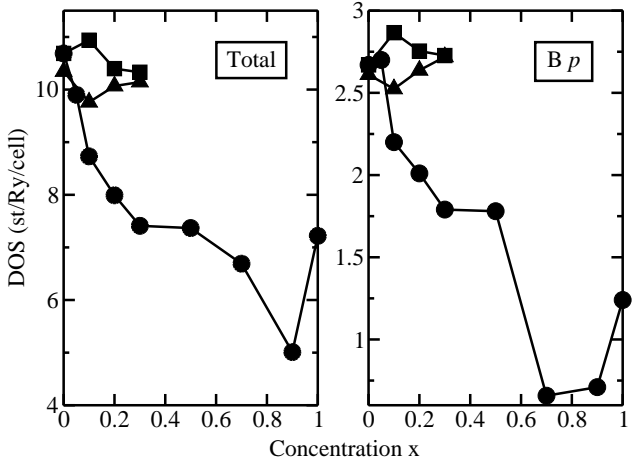,height=7.5cm,angle=-90}
\vspace{0.3cm}
\caption{The calculated total density of states at the Fermi energy (left panel) and the $B$ $p$ 
contribution to the total DOS (right panel) in 
 $Mg_{1-x}Al_xB_2$, 
 $Mg_{1-x}Li_xB_2$ and 
 $Mg_{1-x}Zn_xB_2$ alloys. 
}
\end{figure}

In order to understand the variation of \( T_{c} \) in \( Mg_{1-x}Al_{x}B_{2}, \)
\( Mg_{1-x}Li_{x}B_{2} \) and \( Mg_{1-x}Zn_{x}B_{2} \) alloys as a function
of concentration \( x \), we have analyzed our results in terms of the spectral
functions, the contribution of Boron \( p \)-electrons to the total DOS and
the total DOS. In Fig. 2(a)-(c), we show the spectral functions along \( \Gamma  \) to
\( A \) direction evaluated at $E_F$ in \( Mg_{1-x}Al_{x}B_{2}, \) \( Mg_{1-x}Li_{x}B_{2} \)
and \( Mg_{1-x}Zn_{x}B_{2} \) alloys for \( x=0.1 \) (Fig. 2(a)), \( x=0.3 \)
(Fig.2(b)), and \( x=0.6-1.0 \) (Fig. 2(c)). From Figs. 2(a)-(b) it is clear that
the substitution of \( Al \) in \( MgB_{2} \) leads to creation of more new
states along \( \Gamma  \) to \( A \) direction than the substitution of \( Zn \)
or \( Li. \) Since the hole-like cylindrical Fermi sheet along \( \Gamma  \) to
\( A \) contributes much more to the electron-phonon coupling \cite{cho}, the creation
of new electron states along\( \Gamma  \) to \( A \) direction weakens considerably
the overall coupling constant \( \lambda , \) which, in turn, reduces the \( T_{c} \)
more in \( Mg_{1-x}Al_{x}B_{2} \) than in either \( Mg_{1-x}Zn_{x}B_{2} \)
or \( Mg_{1-x}Li_{x}B_{2} \). Thus, in our opinion, the way \( T_{c} \) changes
in \( MgB_{2} \) upon alloying depends dramatically on the location of the
added/modified \textbf{k}-resolved states on the Fermi surface. 

Having explained the differences in behavior of \( MgB_{2} \) upon alloying
with \( Al, \) \( Li \) and \( Zn, \) we now try to understand the changes
in their properties as a function of concentration \( x. \) In Fig. 3(a)-(b)
we have shown the total DOS at $E_F$ (Fig. 3(a)) and the \( B\, p \) contribution to
the total DOS at $E_F$ (Fig. 3(b)) in \( Mg_{1-x}Al_{x}B_{2}, \) \( Mg_{1-x}Li_{x}B_{2} \)
and \( Mg_{1-x}Zn_{x}B_{2} \) alloys as a function of concentration \( x. \)
We find that as a function of concentration, the variation in \( T_{c} \), as
shown in Fig.1, follows closely the behavior of the total DOS at $E_F$ and in particular
the variation in \( B\, p \) contribution to the total DOS at $E_F$. It is also not
surprising to see that the vanishing of superconductivity in \( Mg_{1-x}Al_{x}B_{2} \)
at \( x\sim 0.7 \) coincides with a very small \( B\, p \) contribution to
the total DOS. 

In Fig. 4(a)-(c) we show the total DOS of \( Mg_{1-x}Al_{x}B_{2} \),  \( Mg_{1-x}Li_{x}B_{2} \)
and \( Mg_{1-x}Zn_{x}B_{2} \) alloys, respectively,  at \( x=0.1. \) In the same plot we also
show the total DOS of \( MgB_{2} \) obtained using the same approach. The overall
downward (upward) movement of the total DOS in \( Mg_{0.9}Al_{0.1}B_{2} \)
(\( Mg_{0.9}Li_{0.1}B_{2} \)) with respect to that of \( MgB_{2} \) is due
to the addition (removal) of electrons. In Fig. 4(c), the peak in the total
DOS at around \( 0.53\, Ry \) below $E_F$ is due to the \( 3d \)
states of \( Zn. \) 
\begin{figure}
\centering
\psfig{file=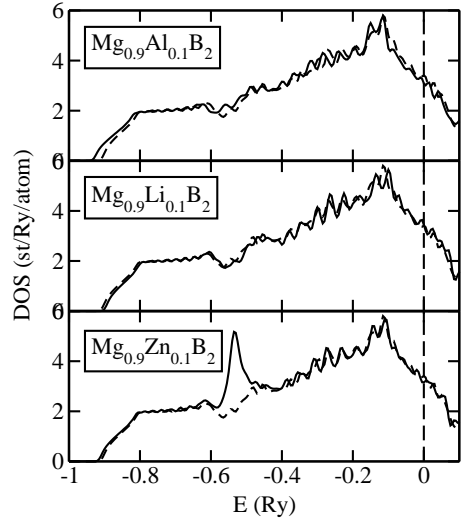,height=7.5cm,angle=-90}
\vspace{0.3cm}
\caption{The calculated total density of states (solid line) for 
 $Mg_{0.9}Al_{0.1}B_2$ (upper panel), 
 $Mg_{0.9}Li_{0.1}B_2$ (middle panel) and
 $Mg_{0.9}Zn_{0.1}B_2$ (lower panel) alloys.  For comparison the total DOS for 
 $MgB_2$ (dashed line) is also shown. The dashed vertical line indicates the Fermi energy.
}
\end{figure}

Finally, we discuss the possibility of superstructure formation at \( x=0.5 \).
In Table I we show the optimized cell parameters as well as the cell-internal
relaxations for \( MgAlB_{4}, \) \( MgLiB_{4} \) and \( MgZnB_{4} \) calculated
using the ABINIT program. We find that the Boron layer shifts significantly
more towards \( Al \) layer in \( MgAlB_{4} \) than towards either \( Li \)
or \( Zn \) layer in \( MgLiB_{4} \) or \( MgZnB_{4} \), respectively. The
shift of \( B \) layer by \( \sim 0.24\, a.u. \) towards \( Al \) layer in \( MgAlB_{4} \)
compares well with the corresponding shift obtained in Ref. \cite{bar}.  However, the shift
of \( B \) layer towards \( Li \) and \( Zn \) layers in \( MgLiB_{4} \)
and \( MgZnB_{4} \) respectively implies that it is not simply due to the extra
positive charge on the impurity layer, as suggested in Ref. \cite{bar} in the case of \( MgAlB_{4}. \)
In Table I we have also listed the calculated ordering energy, \( E_{ord}, \)
which is the difference between the total energies of the ordered \( MgAlB_{4} \),
\( MgLiB_{4} \) and \( MgZnB_{4} \) and the corresponding disordered \( Mg_{0.5}Al_{0.5}B_{2}, \)
\( Mg_{0.5}Li_{0.5}B_{2} \) and \( Mg_{0.5}Zn_{0.5}B_{2} \) alloys, obtained 
using KKR-ASA CPA method. It clearly
shows the possibility of formation of a superstructure in \( Mg_{0.5}Al_{0.5}B_{2} \)
because the fully-relaxed \( MgAlB_{4} \) is lower in energy by \( 12\, mRy/atom \)
in comparison to the disordered \( Mg_{0.5}Al_{0.5}B_{2}. \) However, within the limitations
of our approach, we find that \( Mg_{0.5}Li_{0.5}B_{2} \) and \( Mg_{0.5}Zn_{0.5}B_{2} \)
are unlikely to form superstructures since \( E_{ord} \) is positive in these
two cases. Our results also show that a structure made up of layers consisting
of a random mixing of \( Mg \) and \( Al \) atoms and described by CPA, is higher in
energy than a structure made up of alternate layers of \( Mg \) and \( Al \) atoms \cite{bar}.

\begin{table}
\caption{The calculated lattice constants $a$ and $c/a$, the shift $\delta$ of 
the $B$ layer along the $c$-axis towards the impurity layer, and the ordering 
energy $E_{ord}$ at $x=0.5$.  The lattice constant $a$ and the shift $\delta$ are in atomic units while 
$E_{ord}$ is in ${mRy/atom}$. 
}
{\centering \begin{tabular}{|c|c|c|c|c|}
\hline 
Alloy&
\( a \)&
\( c/a \)&
\( \delta  \)&
\( E_{ord} \)\\
\hline 
\hline 
\( MgAlB_{4} \)&
5.799&
2.242&
0.24&
-12.1\\
\hline 
\( MgLiB_{4} \)&
5.685&
2.287&
0.04&
+4.8\\
\hline 
\( MgZnB_{4} \)&
5.789&
2.254&
0.08&
+1.2\\
\hline 
\end{tabular}\par}
\end{table}
In conclusion, we have shown that in \( Mg_{1-x}Al_{x}B_{2}, \) \( Mg_{1-x}Li_{x}B_{2} \)
and \( Mg_{1-x}Zn_{x}B_{2} \) alloys (i) the way \( T_{c} \) changes depends
on the location of the added/modified \textbf{k}-resolved states on the Fermi
surface, (ii) the variation of \( T_{c} \) as a function of concentration is
dictated by the \( B\, p \) contribution to the total DOS at $E_F$, (iii) at \( x=0.5 \)
a superstructure can form in \( Mg_{1-x}Al_{x}B_{2} \) but not in \( Mg_{1-x}Li_{x}B_{2} \)
or \( Mg_{1-x}Zn_{x}B_{2} \), and (iv) $B$ layer shifts towards the
impurity layer, more for \( Al \) than for \( Li \) or \( Zn. \)

\end{document}